\documentclass[10pt]{iopart}
\usepackage{graphicx}
\usepackage{dcolumn}
\usepackage{bm}
\usepackage{xr}
\usepackage{hyperref}
\usepackage{color}

\begin{document}
\title{Rydberg-induced optical nonlinearities from a cold atomic ensemble trapped inside a cavity}
\author{R Boddeda$^1$, I Usmani$^1$, E Bimbard$^1$, A Grankin$^1$, A~Ourjoumtsev$^2$, E Brion$^3$ and P Grangier$^1$}
\address{$^1$ Laboratoire Charles Fabry, Institut d'Optique Graduate School, CNRS, Universit\'e Paris-Saclay,  91127 Palaiseau, France}
\address{$^2$ Coll\`ege de France,11 Place Marcelin Berthelot, 75005 Paris, France}
\address{$^3$ Laboratoire Aim\'e Cotton, Universit\'e Paris-Sud, ENS Cachan, CNRS, Universit\'e Paris-Saclay, 91405 Orsay Cedex, France}
\date{\today}

\begin{abstract}
We experimentally characterize the optical nonlinear response of a cold atomic medium placed inside an optical cavity, and excited to Rydberg states. The excitation to S and D Rydberg levels is carried out via a two-photon transition in an EIT (electromagnetically induced transparency) configuration, with a weak (red) probe beam on the lower transition, and a strong (blue) coupling beam on the upper transition. The observed optical nonlinearities induced by S states for the probe beam can be explained using a semi-classical model with van der Waals' interactions. For the D states, it appears necessary to take into account a dynamical decay of Rydberg excitations into a long-lived dark state. We show that the measured nonlinearities can be explained by using a Rydberg bubble model with a dynamical decay.
\end{abstract}
\submitto{\jpb}
\maketitle
\ioptwocol

\section{Introduction}
Quantum states of the light are extensively studied for applications to long distance quantum communication and information processing tasks \cite{Caulfi2010,Brien2009,Chang2014}. In order to fully exploit their advantages, a daunting goal is to achieve nonlinearity at the single-photon level, or in other words, to effectively implement strong photon-photon interactions. Such nonlinearity can be obtained through the interaction with a nonlinear auxiliary medium, and many approaches have been considered, e.g. quantum dots \cite{Fushman2008}, cavity QED \cite{Birnbaum2005}, nonlinear waveguides \cite{NonlinearWaveguide2014}, electromagnetically induced transparency (EIT) \cite{Feizpour2015} and more recently Rydberg blockade \cite{Lukin2001,Pritchard2012,Saffman2010}. Thanks to their large dipole moments, Rydberg atoms exhibit strong long-range dipole-dipole interactions. These interactions can shift the excited state of an atom out of resonance with the excitation light, which forbids its excitation, leading to the so-called Rydberg blockade: each Rydberg atom becomes surrounded by a ``blockade sphere'' where no other atom can be excited to a Rydberg state \cite{Lukin2001}. When an atomic ensemble is undergoing EIT through a Rydberg level, the Rydberg blockade affects the EIT conditions and translates into a strong nonlinearity.

Over the past decade there has been considerable progress in achieving strong nonlinearities with Rydberg atoms \cite{Pritchard2010}, which led e.g. to the generation of nonclassical states of light \cite{Peyronel2012,Dudin2012,Firstenberg2013} and single-photon transistors \cite{Baur2014,Tiarks2014,Gorniaczyk2014}. In free space, a basic requirement for few-photons nonlinearity is to obtain a large optical depth per blockade sphere  allowing one photon converted to a Rydberg excitation to strongly modify the cloud's optical response; this implies a large atomic density and can lead to bound states or other undesirable features \cite{Bendkowsky2010}. Instead of increasing the optical depth, an alternative solution is to embed the atomic medium in an optical cavity: even with a low finesse, the increased atom-light coupling provides enhanced nonlinear effects, as we already demonstrated experimentally in the dispersive regime \cite{Parigi2012}. 

The full description of such systems requires solving a many-body problem and constitutes a highly challenging task, in free-space \cite{Gorshkov2011,Gorshkov2013,Sevincli2011} as well as in cavity systems \cite{Grankin2014,Grankin2015}. Experimental investigations can help revealing the most relevant physical features and phenomena, paving the way towards a better understanding and use of such strongly nonlinear systems.  

In this article, we measure Rydberg nonlinearities using a cold atomic ensemble of $^{87}$Rb atoms enclosed in an optical cavity. In particular, we measure the optical transmission of the cavity in the absorptive regime.  In the weak driving limit, we show that the experimental observations agree well with a semi-classical model for S-states. We also give a theoretical explanation for the dynamical effects that we observe in the optical nonlinearity for D-states.

\section{Experimental setup}\label{sec:experiment}
\begin{figure}[hbtp]
\centering
\includegraphics[width=\columnwidth]{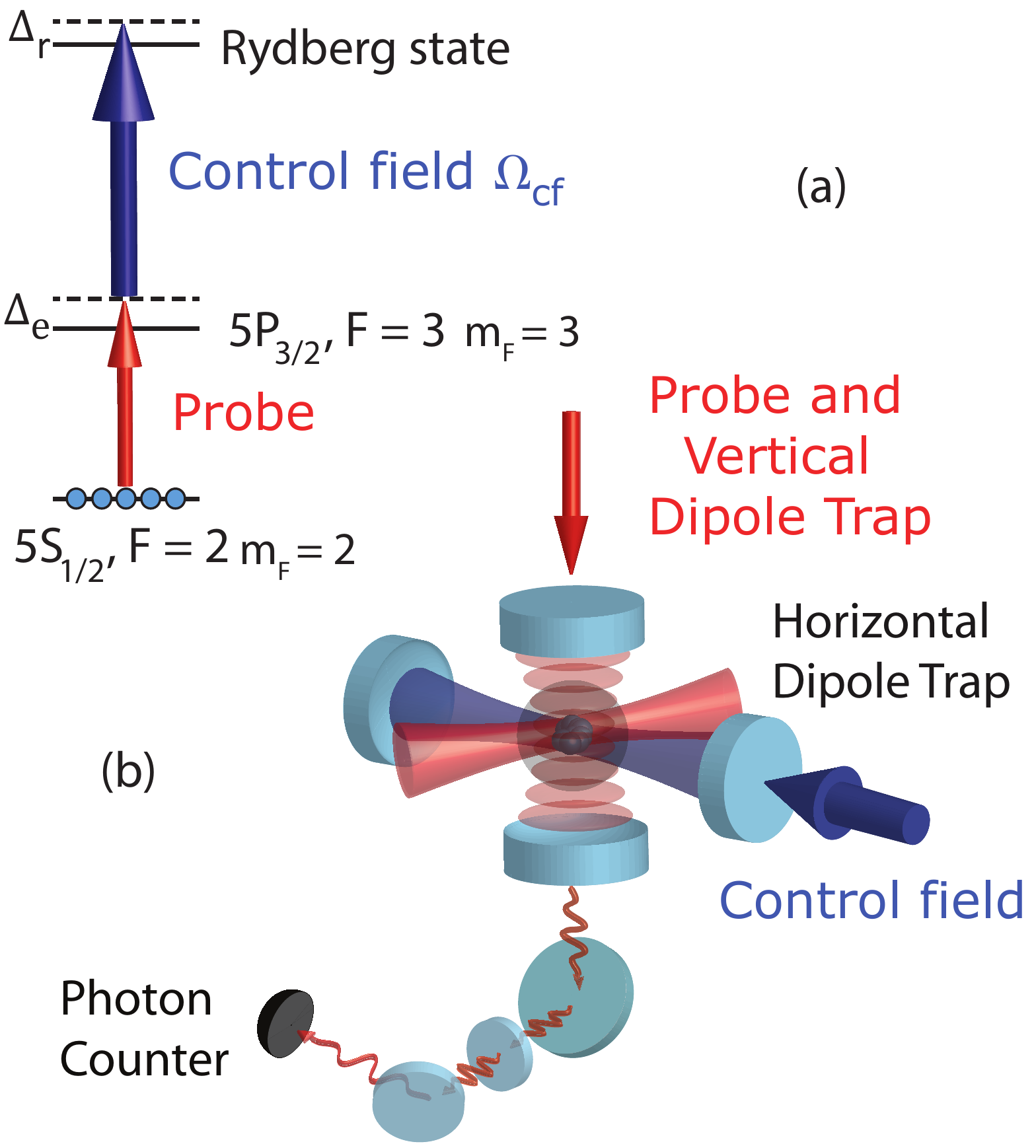}
\caption{(a) Atoms are optically pumped to m$_F$ = 2 Zeeman sublevel of 5S$_{1/2}$ F = 2 in $^{87}$Rb. We simultaneously send two pulses of probe and control field onto the medium and study the properties of transmitted probe beam. (b) An ensemble of $^{87}$Rb atoms are loaded from Magneto-Optical Trap (MOT) to a crossed dipole trap ($w_0 \approx 40$ $\mu m$). The loaded atoms are optically pumped to the desired atomic state and then probed using a two photon transition. The probe beam, transmitted through the cavity, is later analysed using photon counters.}
\label{fig:setup} 
\end{figure}

To measure the non-linear response of a gas driven to Rydberg states we trap a cloud of cold $^{87}$Rb atoms in a cavity. We load $^{87}$Rb atoms into a magneto-optical trap (MOT) from a 2D MOT. Atoms in the MOT are optically cooled down to $\approx$ 50 $\mu$K using polarization gradient cooling in 6~ms. After cooling, the atoms are loaded into a conservative optical dipole potential formed by two crossed dipole traps at 810 nm with a 300 $\mu K$ depth, placed inside a 66 mm long vertical Fabry-Perot cavity with a finesse $\mathcal{F}$ = 120, linewidth $\gamma_c/2\pi$ = 10 MHz and waist $w_0$ = 86 $\mu$m. Then all the atoms in the MOT are depumped to a dark state (5S$_{1/2}$ F = 1),  and only the atoms at the intersection of the dipole traps are selectively repumped to the desired state (5S$_{1/2}$ F = 2, m$_F$ = 2), as shown on Figure \ref{fig:setup}. Once the cloud is prepared, a weak probe is coupled into the cavity driving the transition 5S$_{1/2}$ F = 2, m$_F$ = 2 $\rightarrow$ 5P$_{3/2}$ F = 3, m$_F$ = 3. We scan the detuning of the probe $\Delta_p/2\pi$ from -50 MHz to +50 MHz in 120 $\mu$s. The experimental setup is schematically shown in Figure \ref{fig:setup}b.

To observe a strong EIT signal and to avoid inhomogeneous broadenings, one needs a bright control beam with a uniform intensity, coupling the short-lived state excited by the probe to a long-lived high-lying Rydberg state. We achieved this by using a confocal cavity with a finesse $\mathcal{F}\approx 50$ in a running-wave configuration to enhance the Rabi frequency of the control beam. The smallest beam waist of $90$ $\mu m$, overlapped with the atomic cloud, largely exceeds the $\sigma = 35$ $\mu m$ radius of the latter.

\section{Experimental Results}

In section \ref{sec:EIT}, we look at the transmission of a cavity containing an atomic ensemble undergoing EIT. In this case, we work with a weak probe and a low-lying Rydberg level (n = 37), so that the response of the medium is linear and one can easily obtain an analytical expression for the transmission. In section \ref{sec:NonlinS}, we investigate the nonlinear transmission of our system when the atoms are coupled to a Rydberg S state with a large principal quantum number. We use a theoretical model from \cite{Grankin2015} and a mean field theory approach to describe our experimental observations. Then, in section \ref{sec:NonlinD}, we investigate the nonlinearity of the atoms coupled to the Rydberg D-states. The associated Rabi frequency of the coupling field is expected to be higher, so that one can explore Rydberg levels with higher principal quantum numbers and larger dipole moments. Despite this promising aspect, quantum nonlinearities have never been observed with these states. Our measurement of the nonlinear absorption and the use of a phenomenological model will help to understand the observations.

\subsection{\label{sec:EIT}Intracavity EIT}

\begin{figure}[hbtp]
\centering
\includegraphics[width=\columnwidth]{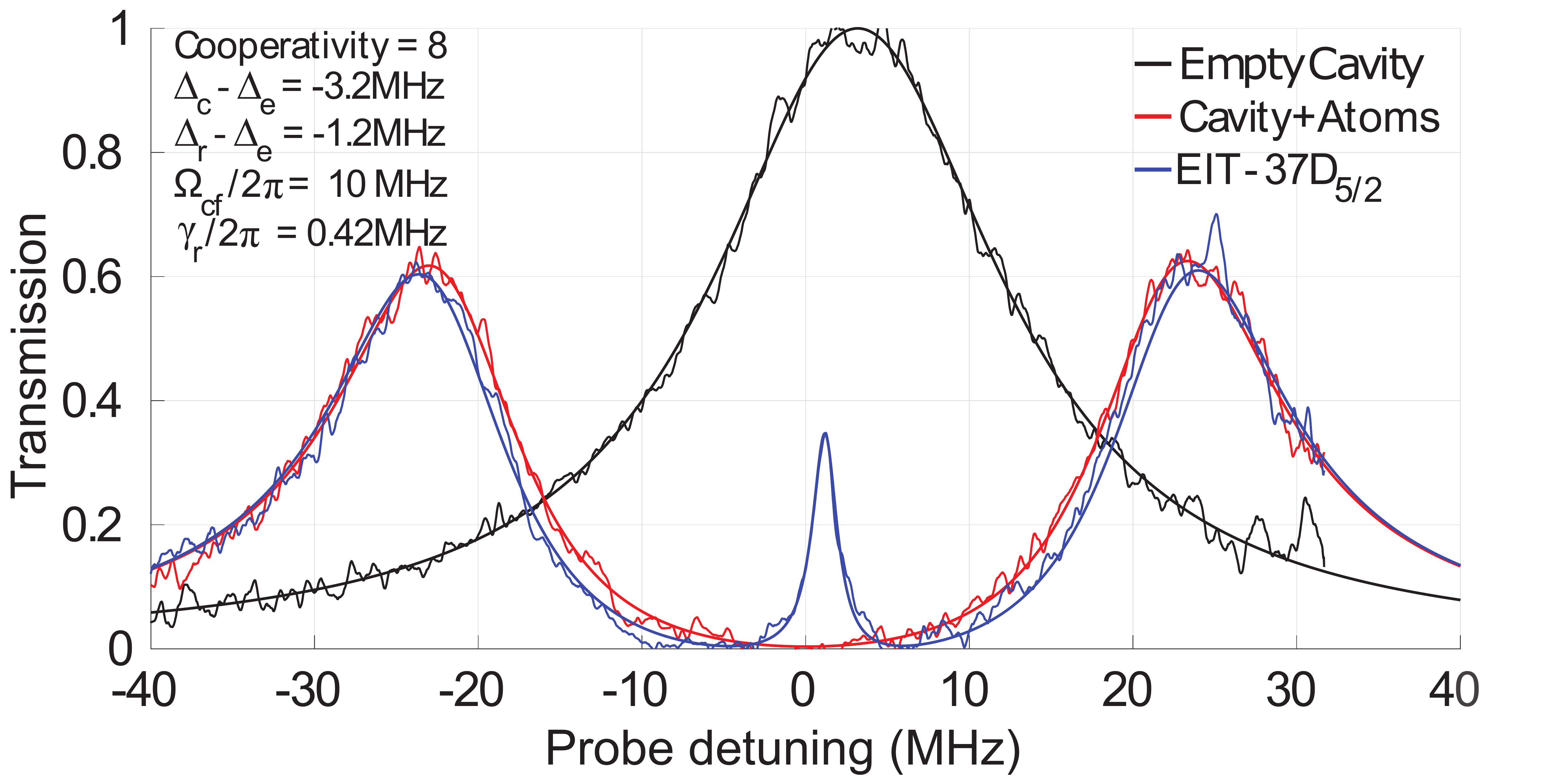}
\caption{Cavity transmission as a function of the probe laser detuning for an empty cavity (black curve), with an intracavity atomic cloud (red curve) and with the presence of a control field resonant with the $5P_{3/2} F=3 \rightarrow 37D_{5/2} F=4 $ transition (blue curve). A fit to the experimental observations allows us to extract various parameters of the system. We note that the centre of the two normal modes doesn't coincide exactly with the empty cavity resonance. This effect, taken into account in the fitting curves, is due to the presence of background atoms in a dark state (5S$_{1/2}$ F = 1) which shifts the cavity line.}
\label{fig:EIT}
\end{figure}

To avoid at first any nonlinear effects arising from a Rydberg blockade, we couple the atoms to a level with a low principal quantum number using a control field resonant with the $5P_{3/2}$ F = 3 $\rightarrow 37D_{5/2}$ F = 4 transition. Since the Rydberg blockade effect is negligible, the response is identical for S and D states. In Figure \ref{fig:EIT}, we show the cavity transmission as a function of the probe detuning. If the cavity is empty, we observe the expected Lorentzian shape of the cavity resonance. In the presence of the atomic ensemble but without any control field, the cavity line is split into two normal modes, a consequence of the atom-cavity coupling. Then, in the presence of the control field which couples the atoms to the 37D Rydberg level, a transparency window is created at the two-photon transition frequency. One can obtain an analytical expression to describe all these cases. The Hamiltonian of the system is described in \ref{AppendixA}. If we don't consider the interaction term, it is analytically solvable and we obtain for the transmission in steady state:

$$ T= \left|\frac{\gamma_c\left(\Delta_e+i\gamma_e-\frac{\Omega_{cf}^2}{4(\Delta_r+i\gamma_r)}\right)}{\left(\Delta_e+i\gamma_e-\frac{\Omega_{cf}^2}{4(\Delta_r+i\gamma_r)}\right)(\Delta_c+i\gamma_c)-2\gamma_c\gamma_eC} \right|^2$$

\noindent where $\Delta_k$ and $\gamma_k$ (for k=e,r,c)  are the detunings and linewidth of the single-photon transition, two-photon transition and the cavity respectively. $C=g^2N/(2\gamma_e\gamma_c)$ is the cooperativity with $g$ the single atom-field coupling and $N$ the number of atoms. We use this expression to analyze the experimental data (Figure \ref{fig:EIT}). This allows us to measure the cavity linewidth ($\gamma_c$), the cooperativity (C), the Rabi frequency of the control field ($\Omega_{cf}$) and the linewidth of the Rydberg level ($\gamma_r$). We note that such intracavity linear EIT has recently been observed with a similar setup \cite{Ningyuan2015}.

\subsection{\label{sec:NonlinS}Resonant nonlinearity with Rydberg S state}
 
\begin{figure}[hbtp]
	\centering
	\includegraphics[width=\columnwidth]{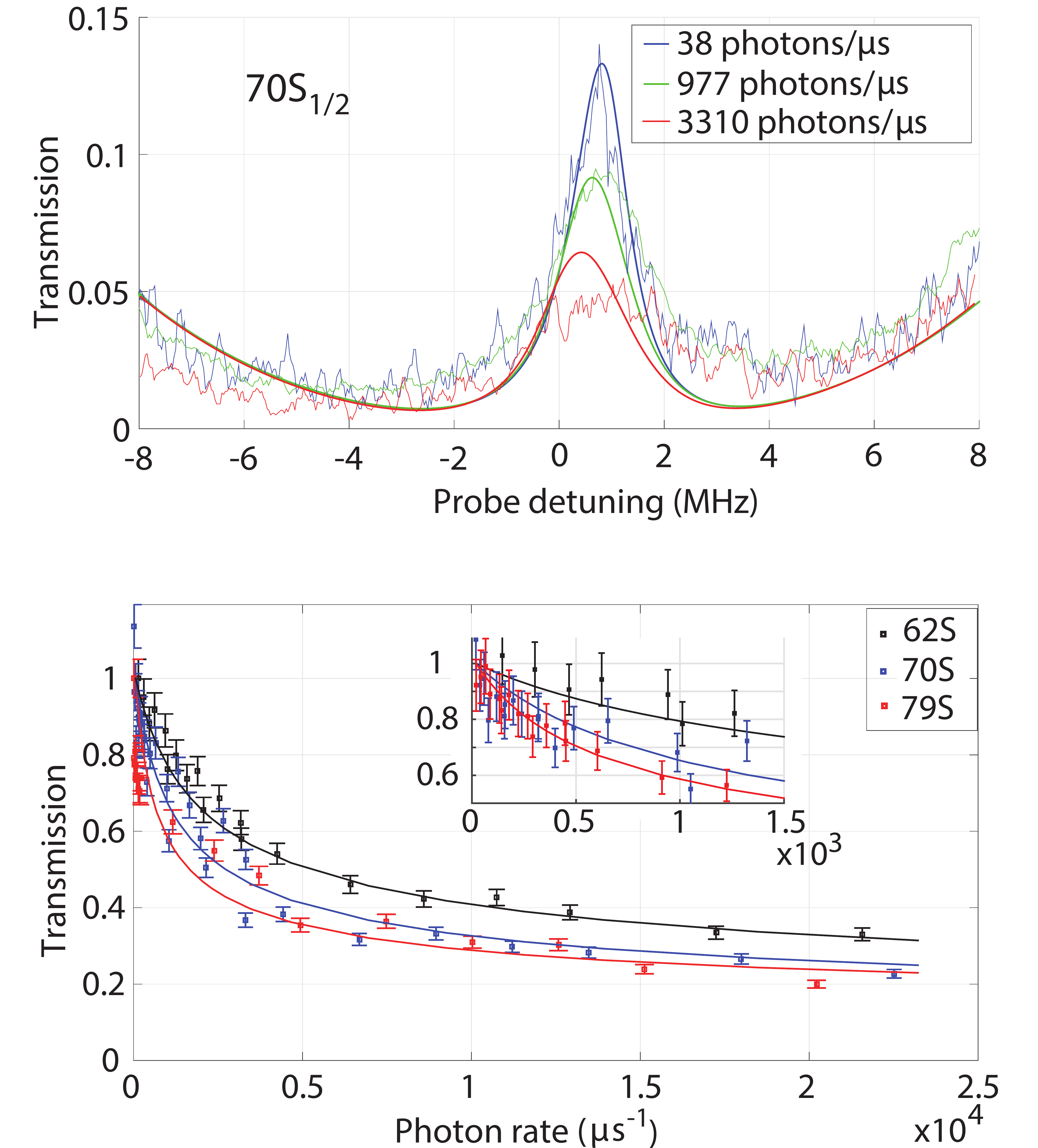}
	\caption{(a) The cavity transmission spectrum is plotted for various probe photon rates for 70S$_{1/2}$ Rydberg state. (b) Cavity transmission at the center of the transparency window as a function of the probe photon rate, for various Rydberg levels. Each curve has been normalized to help their comparison and the inset gives a closer look of the transmission at low photon rates. One can observe the reduction of the transparency with higher photon rate due to the Rydberg blockade effect. The theory is in quite good agreement with the data at any detuning or Rydberg state.}
	\label{fig:nonlinS}
\end{figure}

For S states, the nonlinear effect is induced by a long range and isotropic potential of the form $V(r)=-C_6/r^6$, where $r$ is the interatomic distance and the $C_6$ coefficient for $^{87}$Rb atoms is given by \cite{Singer2005}:
$$C_6 \approx \left(63-267\left(\frac{n}{60}\right)+64\left(\frac{n}{60}\right)^2\right)\left(\frac{n}{60}\right)^{11} \textrm{GHz}\cdot\mu \textrm{m}^6$$
where n is the principal quantum number.
This van der Waals type of interactions creates a blockade effect in EIT conditions \cite{Pritchard2012}, so that the transmission of the medium decreases with increasing probe field strength. This translates, in our case, to a cavity transmission that reduces with the number of photons. We measured this transmission by scanning the probe for various probe photon rates (Figure \ref{fig:nonlinS}a). To bring out more clearly the nonlinear absorption, we take the transmission at zero detuning in Figure \ref{fig:nonlinS}a and plot it as a function of the probe photon rate measured with the same probe power for an empty cavity (Figure \ref{fig:nonlinS}b). This procedure has been repeated for various principal quantum numbers, since we expect a stronger nonlinearity for higher n.

This result can now be compared with the theoretical model from \cite{Grankin2015}. We used a mean field approximation to calculate the expected transmission. More details can be found in \ref{AppendixA}. The experimental parameters such as the cooperativity, Rydberg state linewidth or the Rabi frequency of control field have been measured (section \ref{sec:EIT}), leaving no free parameters in the model. The experimental data appears to be in a relatively good agreement with the theory for any probe detuning or Rydberg level (Figure \ref{fig:nonlinS}), at least for small photon rate, i.e. as long as the fraction of blockaded atoms remains small. This is the first experimental test of the model provided in \cite{Grankin2015} and it appeared to be reliable to predict the nonlinear transmission of the system. We can therefore expect to see, as it is predicted in the model, anti-bunching of the light when the nonlinearity is strong enough. For that, we would need the cooperativity per blockade sphere to be larger than one. In our experiment, however, the maximum value we could obtain was 0.1 and 0.2 for the 79S and 92D states respectively. 

\subsection{\label{sec:NonlinD}Dephasing in Rydberg D-States}

\begin{figure}[hbtp]
\centering
\includegraphics[width=\columnwidth]{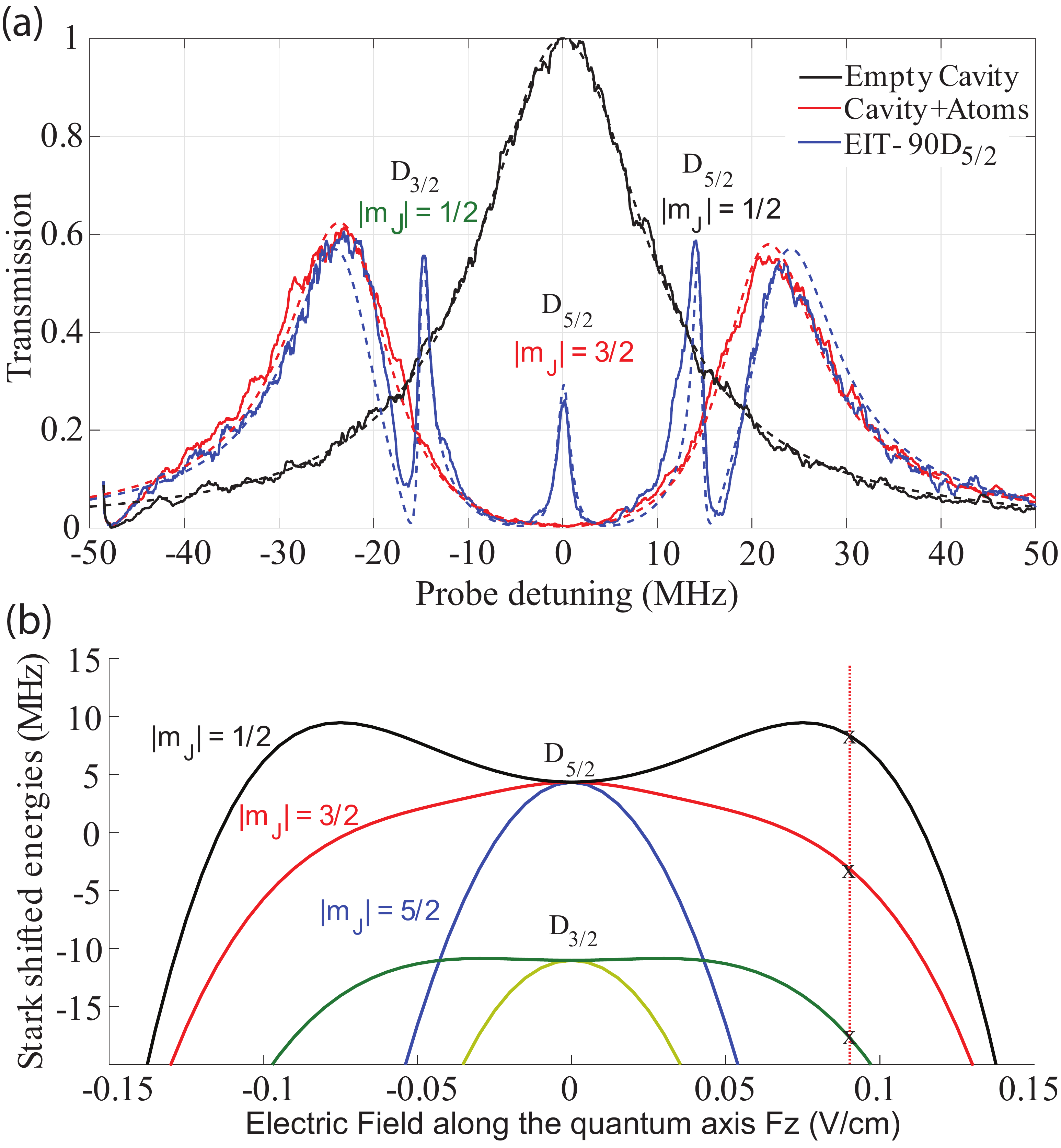}
\caption{(a) A stray electric field lifts the degeneracy for the Zeeman sublevels of the 90D states, which creates a transparency window for each transition. (b) Theoretical detuning of the degenerate levels with respect to the electric field in quantization axis. By comparing it to the measured splittings in (a), we can estimate the stray electric field in the quantization axis to be about 90 mV/cm.}
\label{fig:starkshift}
\end{figure}

Compared to S states, Rydberg D states have larger dipole couplings to the intermediate $5P_{3/2}$ state, which allows one to reach states with higher principal quantum numbers for the same coupling light intensity, and thus to get higher nonlinearities due to the associated increase of the blockade volume. However, no quantum effects such as anti-bunching have been observed yet for D states even though several experiments have been conducted to observe a nonlinearity in such systems \cite{Pritchard2011, Hofferberth2015}. One must note two differences with S states which might complicate the blockade effect. Primarily, the Rydberg interactions for D states are anisotropic, they can even vanish for some orientations \cite{Saffman2010}. This angular dependance has been directly measured for two atoms in the 82D$_{3/2}$ state \cite{Barredo2014}. In our model, we considered however an isotropic interaction \cite{Stanojevic2013}: $C_6=45\cdot\left( n/56 \right)^{11} $GHz$\cdot\mu m^6$, averaging the potential in all directions. This simplification allows one to keep analytical calculations tractable, but it prevents one from correctly modeling beyond-mean-field effects related to the anisotropy of the interactions. Secondly, the excitation to D states involves a large number of Zeeman sublevels which are highly sensitive  to the electric field, especially at high principal quantum numbers. In our case, a small stray electric field was sufficient to lift the degeneracy as shown in a spectroscopic scan for a high Rydberg level (Figure \ref{fig:starkshift}a). The observed splittings are in good agreement with a simple Stark shift model \cite{Beguin2013}, see Figure \ref{fig:starkshift}b.

\begin{figure}[hbtp]
\centering
\includegraphics[width=\columnwidth]{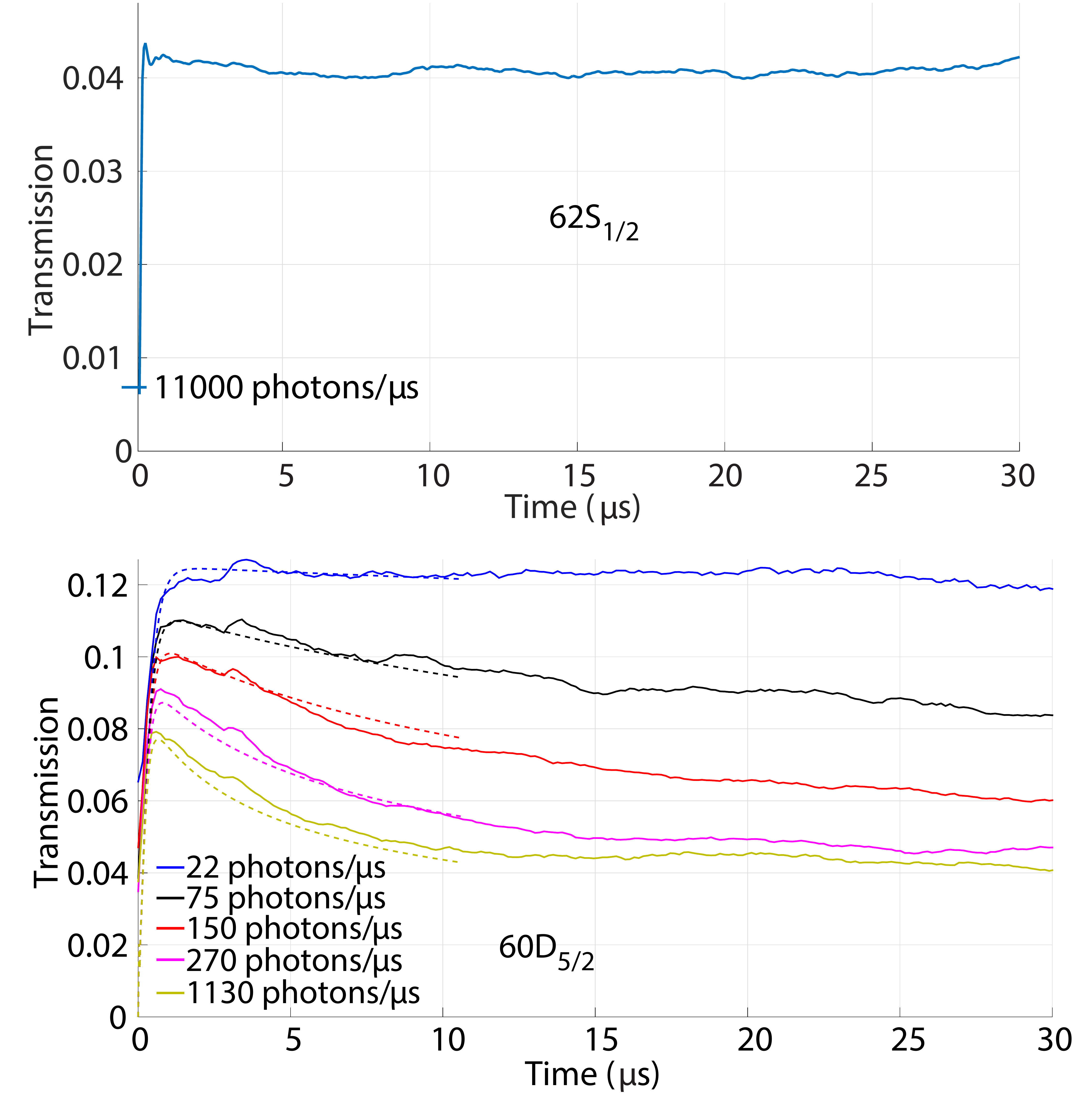}
\caption{Cavity transmission in EIT conditions  (probe and control fields on resonance) for S(a) and D(b) states. In addition to a decrease in transmission with the probe photon rate due to the Rydberg blockade, we observe a transmission decay over time for D states. This can be very well fitted with a model including a possible decay of atoms into other Rydberg states that are uncoupled to the control field.}
\label{fig:transient}
\end{figure}

By exciting atoms to D states (keeping the probe and control field on resonance), we observe a transient decrease of the transmission during $\approx$10 $\mu$s (Figure \ref{fig:transient}(b)). This dynamical behaviour was also observed in \cite{Hofferberth2015} and is neither present for S states (Figure \ref{fig:transient}(a)), nor predicted by the steady-state model we used in section \ref{sec:NonlinS}. In addition, we have also observed a decrease in transmission even after switching off the excitation lasers for 50 $\mu s$ and probing it with a weak probe beam. The decay time of 10 $\mu s$ is far from any timescales present in our steady-state model. To explain this, we consider that, because of the anisotropy of D states and their multilevel structure, a Rydberg atom can decay into another level that is uncoupled from the control field, a dark Rydberg level. This process creates long-lived blockading Rydberg atoms which would decrease the total transmission.  This transient decay, slower than all the timescales predicted by our simple blockade model, is thus the sign of the creation of long-lived blockading atoms. In \cite{Hofferberth2015}, the Zeeman sublevels were degenerate and the decay was also explained by the evolution of Rydberg atoms into such levels. It is therefore interesting to notice that the dynamical behaviour is still present in our case, even if an electric field has lifted the degeneracy. 

To account for this process, we included a decay term $\xi$ in the bubble model (see \ref{AppendixB}), to account for the decay rate into ``dark'' Rydberg states. This gives us a free parameter to fit the curves in Figure \ref{fig:transient}(b). The good agreement with the data - for short times, i.e. when the medium is not saturated with blockade spheres - indicates that it is a good phenomenological model.

\begin{figure}[hbtp]
\centering
\includegraphics[width=\columnwidth]{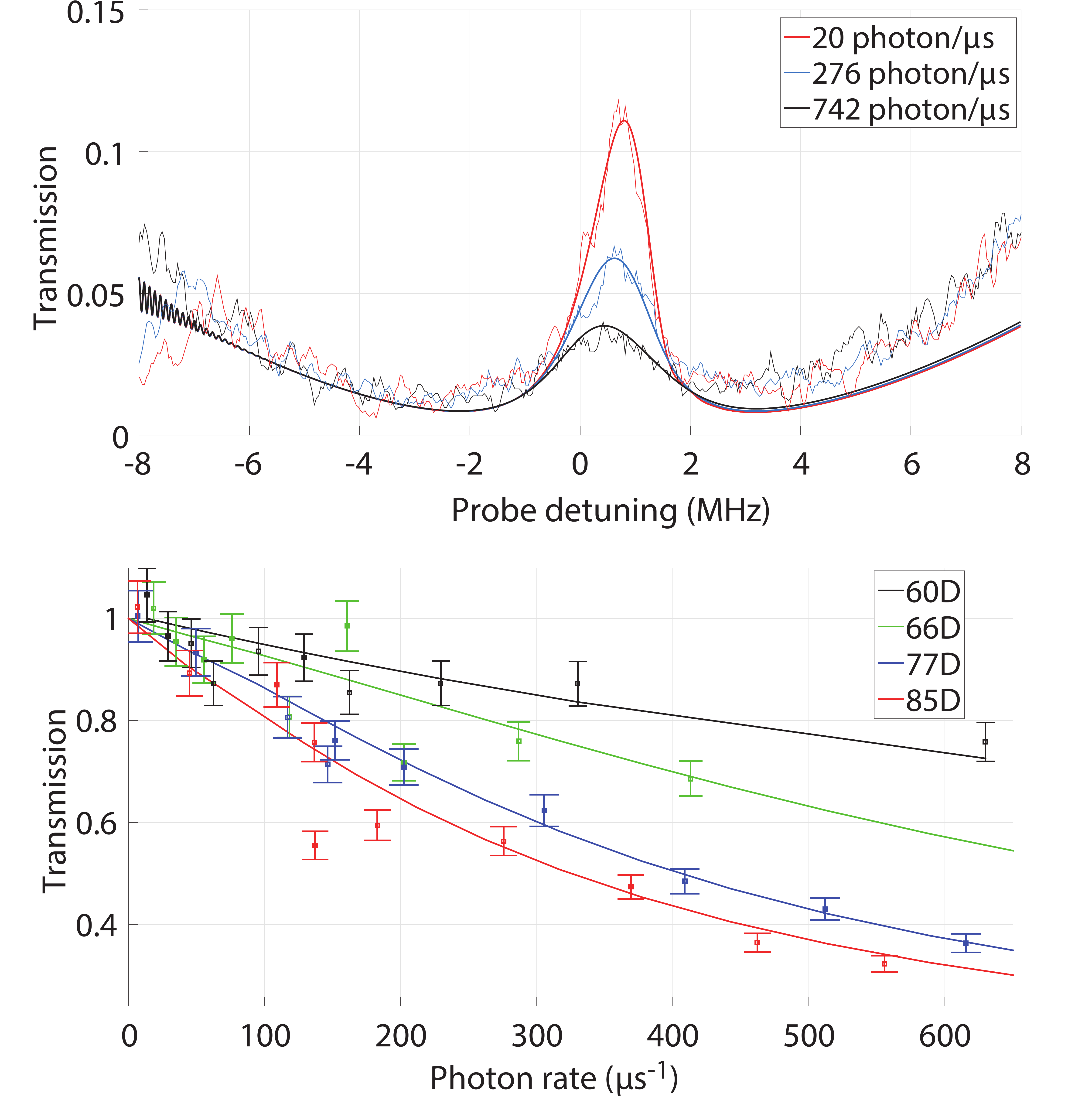}
\caption{(a) The cavity transmission spectrum is plotted for various photon rates for 85D$_{5/2}$ Rydberg state (b) Cavity transmission at the center of the transparency window as a function of the photon rate, for various Rydberg levels. The model including long-lived Rydberg atoms agrees well with the observed transmission data.}
\label{fig:nonlind}
\end{figure}

Even while scanning the probe rapidly around the two-photon resonance, we observed a nonlinear transmission similar to S states. The nonlinearity appears to be stronger, although the $C_6$ coefficient for D states are of the same order. The reason is that the transmission is affected by the creation of long-lived blockading atoms during the scan. Including the decay to dark Rydberg levels into our model allows us to fit the data with good accuracy (Figure \ref{fig:nonlind}) for all detunings and various principal quantum numbers. We obtained a decay parameter for each principal quantum number: $\xi=\{1.8\pm1.7, 2.2\pm1.7, 2.3\pm1.7, 1.1\pm0.66\}$ MHz for $n=\{60, 66, 77, 85\}$, where the error bars contain a 95\% confidence interval. While this parameter allows to fit the data with good accuracy, the large incertitude doesn't permit to conclude about it's dependence on the principal quantum number. 

We believe that the creation of such long-lived Rydberg atoms is detrimental to the production of non-classical light states via the blockade-induced optical non-linearity of the atomic cloud. Indeed, after the medium is saturated with long-lived blockading atoms, the medium has a low transmission, but it has a linear response to an incoming field. This may explain why effects such as photonic antibunching have never been observed in experiments where atoms were driven to Rydberg D states. To use such transitions for quantum optics experiments, this decay mechanism must be circumvented.

\section{Conclusion}
In this article, we described cavity-enhanced optical nonlinearity measurements in a Rydberg atomic ensemble. We proposed a semi-classical model to explain our observations for atoms driven to Rydberg S states. We also modeled the dynamical dephasing effects observed for Rydberg D states by a decay to long-lived dark states. 

\section*{Acknowledgements}
This work is supported by the European Union grants 
SIQS (FET No. 600645) and RySQ (FET No. 640378), and by the Chaire SAFRAN--IOGS Photonique Ultime. 

\appendix

\section{Semi-classical model} \label{AppendixA}
We use here the results from \cite{Grankin2015}, which describes well our intracavity Rydberg medium, to obtain the nonlinear transmission of the system in the mean field approximation.

We use the same notations as in \cite{Grankin2015}. The system consists of a cloud of atoms with a three-level ladder structure with ground, intermediate and Rydberg states are denoted by g, e and r respectively. The decay rates of intermediate and Rydberg states are denoted by $\gamma_e$ and $\gamma_r$ respectively. The respective detunings of the probe field with respect to the atomic transition ($ g \rightarrow e$) and the cavity resonance are denoted by $\Delta_e \equiv \omega_p - \omega_{eg}$ and $\Delta_c\equiv \omega_p - \omega_c$. The feeding of the cavity mode by the probe field is denoted by $\alpha$ and the annihilation operator of the mode of the cavity is denoted by $a$. We introduce atomic operators that are denoted by $\hat{\sigma}_{kl}^{(i)}=|k^{(i)}\rangle\langle l^{(i)}|$ for each atom $i$. The Hamiltonian considered is:

\begin{eqnarray} 
H &= -\Delta_e \sum_i \hat{\sigma}^{(i)}_{ee} - \Delta_r \sum_i \hat{\sigma}^{(i)}_{rr} - \Delta_c \hat{a}^{\dagger} \hat{a} + \alpha \left(\hat{a} + \hat{a}^{\dagger} \right) \nonumber \\
&   + g \sum_i \left( \hat{a} \hat{\sigma}^{(i)}_{eg} + \hat{a}^{\dagger} \hat{\sigma}^{(i)}_{ge} \right) + \frac{\Omega_{cf}}{2} \sum \left( \hat{\sigma}^{(i)}_{er} + \hat{\sigma}^{(i)}_{re} \right) \nonumber \\ 
 & + \frac{1}{2} \sum_{i \neq j} \kappa_{i,j} \hat{\sigma}^{(i)}_{rr} \hat{\sigma}^{(j)}_{rr} 
\end{eqnarray}

\noindent where $g$ and $\Omega_{cf}$ denotes the single atom-field coupling and the Rabi frequency of the control field respectively. The interaction term is restricted to the symmetric subspace, where states are invariant under the permutation of two particles and $\kappa_{i,j}$ is the average pair interaction.

It was shown in \cite{Grankin2015} that the expectation values $\left\langle \hat{a} \right\rangle$, $\langle \hat{b} \rangle$, $\left\langle \hat{c} \right\rangle$ for the cavity mode annihilation operator and the following collective operators

\begin{eqnarray*}
\hat{b} &= \frac{1}{\sqrt{N}} \sum_i \hat{\sigma}^{(i)}_{ge} \\
\hat{c} &= \frac{1}{\sqrt{N}} \sum_i \hat{\sigma}^{(i)}_{gr}
\end{eqnarray*}

\noindent are governed by the following dynamical equations:

\begin{eqnarray*}
\frac{d}{dt}\left\langle\hat{a}\right\rangle &= (i\Delta_c-\gamma_c) \left\langle\hat{a}\right\rangle - i g\sqrt{N} \langle\hat{b}\rangle - i \alpha \\
\frac{d}{dt}\langle\hat{b}\rangle &= (i\Delta_e-\gamma_{ge}) \langle\hat{b}\rangle - i\sqrt{N}g\left\langle\hat{a}\right\rangle - i\frac{\Omega_{cf}}{2} \left\langle\hat{c}\right\rangle \\
\frac{d}{dt}\left\langle\hat{c}\right\rangle &= (i\Delta_r-\gamma_{gr}) \left\langle\hat{c}\right\rangle - i\frac{\Omega_{cf}}{2} \langle\hat{b}\rangle - i \kappa \langle c^{\dagger} cc \rangle
\end{eqnarray*}

The complex constant $\kappa$ characterizes the effect of dipole-dipole Rydberg interactions, and was found to be 
\begin{eqnarray*}
\kappa = -2 \Biggl(\frac{V_b}{V-V_b}\Biggl)\Biggl(\frac{\Omega_{cf}^2}{4(D_e+ D_r-\frac{\Omega_{cf}^2}{4(D_e)})} -(D_r)\Biggl) \\
V_b =  \frac{\sqrt{2}\pi^2}{3} \sqrt{\frac{C_6}{D_e-\frac{\Omega_{cf}^2}{4(D_e+D_r-\frac{\Omega_{cf}^2}{4D_e})}}} 
\end{eqnarray*}
where $D_k \equiv \Delta_k + i\gamma_{gk}$ for $k = e,r$, $V_b$ is the blockaded volume of a single blockade and $V$ is the total volume of the cloud.

In steady state conditions and mean field approximation, one can derive the following equations
\begin{eqnarray*}
\langle\hat{a}\rangle &= -\frac{ \langle\hat{b}\rangle g\sqrt{N}}{\Delta_c+i\gamma_c} - \frac{\alpha}{\Delta_c+i\gamma_c} \\
\langle\hat{b}\rangle&= -\frac{ \langle\hat{a}\rangle g\sqrt{N}}{\Delta_e+i\gamma_{ge}} - \frac{\langle\hat{c}\rangle \Omega_{cf}}{2(\Delta_e+i\gamma_{ge})} + \frac{g\sqrt{N}\langle\hat{a}\rangle}{\Delta_e+i\gamma_{ge}} \\
\langle\hat{c}\rangle&= -\frac{\langle\hat{b}\rangle \Omega_{cf}}{2(\Delta_r+i\gamma_{gr})} - \frac{\langle\hat{c}\rangle^2 \kappa\bar{\langle\hat{c}\rangle}}{\Delta_r+i\gamma_{gr}} 
\end{eqnarray*}

The average interaction term $\kappa$ can be calculated in the large volume approximation (V $\gg V_b$). The cavity transmission is then easily calculated from the cavity field $\langle\hat{a}\rangle$. 
\begin{eqnarray*}
T = \frac{\gamma_c^2.|\langle\hat{a}\rangle|^2}{\alpha^2}
\end{eqnarray*}

We compared this theoretical result to the measurement of a nonlinear transmission for atoms coupled to a Rydberg S-state (section \ref{sec:NonlinS}).

\section{Rydberg Bubble model} \label{AppendixB}

In this Appendix we give details about the phenomenological model
we used to account for the dynamical behavior of the cavity transmission,
in the case of $D$-Rydberg-state-excited samples.

We assume that the Rydberg blockade phenomenon effectively splits
the atomic sample into independent and equivalent ``bubbles''
which can at most accommodate for one Rydberg excitation. Accordingly,
the two-photon transition towards the Rydberg level $\left|r\right\rangle $
essentially couples the two collective symmetric states 
\begin{eqnarray*}
\left|G\right\rangle  & \equiv & \left|g\cdots g\right\rangle \\
\left|R\right\rangle  & \equiv & \frac{1}{\sqrt{n_{b}}}\sum_{i=1}^{n_{b}}\sigma_{rg}^{\left(i\right)}\left|G\right\rangle =\frac{1}{\sqrt{n_{b}}}\left(\left|rg\cdots g\right\rangle +\cdots\left|g\cdots gr\right\rangle \right)
\end{eqnarray*}
where $n_{b}$ denotes the number of atoms in a Rydberg bubble, and
the corresponding lowering operator is the Pauli-like matrix $\sigma_{GR}\equiv\left|G\right\rangle \left\langle R\right|$
(we also define $\sigma_{RR}\equiv\left|R\right\rangle \left\langle R\right|$
and $\sigma_{RG}\equiv\left|R\right\rangle \left\langle G\right|$
). Note that $n_{b}$ can be evaluated by $n_{b}=N\times\left|\frac{V_{b}}{V}\right|$,
where $N$ is the total number of atoms in the sample, $V$ is the
total volume of the sample and $V_{b}$ is the volume of a Rydberg
bubble whose expression was given in Appendix A.

In a Rydberg bubble, the intermediate state can, by contrast, be arbitrarily
populated; we, however, further assume that we remain in the low
excitation regime (corresponding to moderate cavity feeding rates)
so that the transition to the intermediate state is never saturated.
In this approximation scheme, the collective lowering operator
\[
\beta\equiv\frac{1}{\sqrt{n_{b}}}\sum_{i=1}^{n_{b}}\sigma_{ge}^{\left(i\right)}
\]
can be considered bosonic, \emph{i.e.}$\left[\beta,\beta^{\dagger}\right]\approx1$.

Here, in order to account for the dynamical behavior observed experimentally,
we moreover introduce an extra Rydberg state, denoted by $\left|s\right\rangle $,
to which the state $\left|r\right\rangle $ decays: this implies that,
in a bubble, the collective states $\left|R\right\rangle $ and $\left|S\right\rangle \equiv\frac{1}{\sqrt{n_{b}}}\left(\left|sg\cdots g\right\rangle +\cdots \left|g\cdots gs\right\rangle \right)$
are coupled by a Lindblad-like operator.

To simplify the treatment, we furthermore assume the cavity mode to
be classical, that is we replace $a$ by its expectation value $\left\langle a\right\rangle $
whose time evolution is ruled by the equation
\[
\frac{d}{dt}\left\langle a\right\rangle =\mbox{i}\left(\Delta_{c}+\mbox{i}\gamma_{c}\right)\left\langle a\right\rangle -\mbox{i}\left(\frac{N}{n_{b}}\right)g\sqrt{n_{b}}\left\langle \beta\right\rangle -\mbox{i}\alpha
\]
Note that the second term of this equation arises from the coupling
of the cavity mode with the ensemble of $\left(\frac{N}{n_{b}}\right)$
Rydberg bubbles with the magnified coupling strength $g\sqrt{n_{b}}$.
The first term accounts for the detuning and decay of the cavity,
while the last one results from the feeding by the probe field. In
this semi-classical approximation, bubbles do not entangle with the
cavity mode and therefore cannot get entangled with each other: the
atomic sample can hence be described by the tensor product density
matrix $\rho\otimes\cdots\otimes\rho$ where $\rho$ is the density
matrix of any of the bubbles (they are all equivalent). The semi-classical
dynamical equation for $\rho$ now writes $\frac{d}{dt}\rho=-\mbox{i}\left[H,\rho\right]+\mathcal{D}_{l}\left(\rho\right)+\mathcal{D}_{nl}\left(\rho\right)$
where
\begin{eqnarray*}
H & = & -\Delta_{r}\left|R\right\rangle \left\langle R\right|-\Delta_{e}\beta^{\dagger}\beta\\
 &  & +\left\{ \left(\frac{\Omega_{cf}}{2}\sigma_{RG}+g\sqrt{n_{b}}\left\langle a\right\rangle ^{*}\right)\beta+\mbox{h.c.}\right\} \\
\mathcal{D}_{l}\left(\rho\right) & = & \gamma_{e}\left(2\beta\rho\beta^{\dagger}-\beta^{\dagger}\beta\rho-\rho\beta^{\dagger}\beta\right)\\
 &  & +\gamma_{r}\left(2\sigma_{GR}\rho\sigma_{RG}-\sigma_{RR}\rho-\rho\sigma_{RR}\right)\\
 &  & +\gamma_{s}\left(2\sigma_{GS}\rho\sigma_{SG}-\sigma_{SS}\rho-\rho\sigma_{SS}\right)\\
\mathcal{D}_{nl}\left(\rho\right) & = & \xi\left\langle \sigma_{RR}\right\rangle \left(2\sigma_{SR}\rho\sigma_{RS}-\sigma_{RR}\rho-\rho\sigma_{RR}\right)
\end{eqnarray*}
Note that the phenomenological extra nonlinear decay $\mathcal{D}_{nl}\left(\rho\right)$ we introduced is time dependent through $\left\langle \sigma_{RR}\right\rangle \left(t\right)$; its rate is moreover governed by the \emph{ad hoc }free parameter $\xi$, whose value can be tuned so as to reproduce the experimental
results.

\newpage
\bibliography{CavityRydberg}

\end{document}